\documentclass[12pt]{article}
\begin{document}
\begin{large}
\title{\bf A Newtonian Hidden Variable Theory.}
\end{large}
\author{Bruno Galvan \footnote{Electronic address: b.galvan@virgilio.it}\\ \small Loc. Melta 40, 38014 Trento, Italy.}
\date{\small June 2004}
\maketitle
\begin{abstract}
A new hidden variable theory is proposed, according to which particles follows definite trajectories, as in Bohmian Mechanics or Nelson's stochastic mechanics; in the new theory, however, the trajectories are classical, i.e. Newtonian. This result is obtained by developing the following concepts: (i) the essential elements of a hidden variable theory are a set of trajectories and a measure defined on it; the Newtonian HCT will be defined by giving these two elements. (ii) The universal wave function has a tree structure, whose branches are generated by the measurement processes and are spatially disjoined. (iii) The branches have a classical structure, i.e. classical paths go along them; this property derives from the fact that the paths close to the classical ones give the main contribution to the Feynman propagator. (iv) Classical trajectories can give rise to quantum phenomena, like for instance the interference phenomena of the two-slit experiment, by violating the so called Independence Assumption, which is always implicitely made in the conceptual analysis of these phenomena.
\end{abstract}
\vspace{5 mm}

\section{Introduction}
Hidden-Configuration Theories \cite{family} or Configuration-Space Theories \cite{nelson revisited} (hereafter HCT) are hidden variables theories in which all the particles of the universe have a definite position at every time. The two major examples of HCT are Bohmian Mechanics \cite{bohm}\cite{holland}\cite{undivided} and Nelson's stochastic mechanics \cite{nelson1}\cite{nelson2}. The possibility for a HCT to reproduce the same experimental results of orthodox quantum mechanics rests on the assumption, which has been made by many authors, that all the measurements can be reduced to position measurements \cite{feynman} \cite{bell} \cite{goldstein}. Thus, in order to reproduce the quantum mechanical results, the equation
\begin{equation}
\rho(t,x)=|\Psi(t,x)|^2
\label{1}
\end{equation}
is considered a necessary condition, where $\rho(x,t)$ is the density of the trajectories of the particles and $\Psi(x,t)$ is the quantum wave function.

Equation (\ref{1}) must hold for the universe as well as for any of its subsystems. This is well explained in \cite{absolute} for the case of Bohmian Mechanics, and one of the main result of that paper is to show that (\ref{1}) holds for every subsystem when it holds for the whole universe. In the present paper equation (\ref{1}) and the related concepts will be studied with respect to the whole universe.

Bohmian Mechanics and Nelson's stochastic mechanics are only two of the possible HCT which give rise to (\ref{1}), and there are several studies to find more general theories which give rise to (\ref{1}) \cite{bohmian revisited}\cite{nelson revisited}.

\vspace{3 mm}
In this paper a new HCT is proposed, whose trajectories are classical. The plan of the paper is the following: in section 2 it is shown that a generic HCT can be expressed in the mathematical language of stochastic processes, whose essential elements are a set of trajectories and a measure defined on it. The Newtonian HCT will be defined by giving these two elements. In section 3 it is shown that the universal wave function has a tree structure, whose branches have disjoined support; the branches are generated by the measurement processes. In section 4 it is shown that the branches have a classical structure, i.e. classical trajectories go along them. This propriety derives from the fact that the paths which are close to the classical paths give the main contribution to the Feynman propagator. In section 5 the Newtonian HCT is defined, and it is shown that it satisfy a macroscopic version of condition (\ref{1}); the new condition is however sufficient to guarantee the empirical equivalence between the theory and quantum mechanics. In section 6 the two-slit experiment is discussed, and the interference phenomena are explained in the context of a Newtonian HCT by assuming the violation of the so called Independence Assumption, which is always implicitely made in the analysis of conceptual experiments like the two-slit experiment or the EPR paradox.

\section{Hidden-configuration theories as stochastic processes}

Let $(\Lambda,{\cal F},\mu)$ be a probability space, $(M, {\cal B})$ a measurable space and $T$ an index set, for instance $R$ o $R^+$. A stochastic process with index set $T$ and state space $M$ is a collection of random variables $\{z_t:\Lambda\rightarrow M\}_{t\in T}$. Given $\lambda\in\Lambda$, the map $z_\lambda:T\rightarrow M$, defined by $ z_\lambda(t):=z_t(\lambda)$, is said a sample path. Moreover, let $\mu_t$ denote the measure $z_t \mu$ on $M$, (i.e. $\mu_t(\Delta)=\mu[z^{-1}_t(\Delta)]$, $\Delta\in{\cal B}$).

An equivalent way to define a stochastic process is to define a set $\Lambda$ of trajectories $\lambda:T\rightarrow M$ (the set of sample paths), and to endowe it with a $\sigma$-algebra ${\cal F}$ and a probability measure $\mu$, such in a way that the map $z_t:\Lambda\rightarrow M$, defined by $z_t(\lambda):=\lambda(t)$, is measurable for all $t\in T$.

\vspace{3mm}
Both Bohmian Mechanics and Nelson's stochastic mechanics can be expressed in the mathematical language of stochastic processes. As to Bohmian mechanics, the state of an N-particles universe at the time $t$ is represented by the pair $(x(t), \Psi(t))$, where $x(t)\in R^{3N}$ is the position of the particles and $\Psi(t) \in L^2(R^{3N})$ is the quantum state. The quantum state evolves in time according to the equation 
\begin{equation}
\Psi(t)=U(t)\Psi_0,
\end{equation} \label{3}
where $U(t)$ is the unitary time evolution operator, and $\Psi_0$ is the normalized quantum state of the universe at the time $t=0$, which is considered the initial time of the universe. The position $x(t)$ evolves according to the guidance equation
\begin{equation}
\frac{dx}{dt}=\hbar m^{-1}\Im \frac{\nabla\Psi}{\Psi},
\end{equation}
where m is a matrix. The position of the particles at the initial time is random, distributed according to $|\Psi_0(x)|^2$.

In order to express Bohmian Mechanics as a stochastic process, let us define: (1) the set  $\Lambda$ as the set of trajectories $\lambda:R^+\rightarrow R^{3N}$ which satisfy the quidance equation; note that, with this definition, for all $t$ the map $z_t:\lambda \mapsto \lambda(t)$ is a bijection between $\Lambda$ and $R^{3N}$. (2) The $\sigma$-algebra ${\cal F}$ by ${\cal F}:=z^{-1}_t[{\cal B}]$, where ${\cal B}$ is the Borel $\sigma$-algebra on $R^{3N}$ (this definition does not depend on the time). (3) The measure $\mu$ by $\mu(\Sigma):=\langle\Psi(t)|E[z_t(\Sigma)]|\Psi(t)\rangle$, where $\Sigma\in {\cal F}$ and $E[z_t(\Sigma)]$ is the projection on the spatial region $ z_t(\Sigma)$; due to the equivariance propriety of Bohmian Mechanics, also this definition does not depend on the time.

With these definition, Bohmian Mechanics can be considered as a stochastic process with index set $ R^+$ and state space $R^{3N}$. Note that for Bohmian Mechanics the following equation
\begin{equation}
\mu_t(\Delta)=\langle\Psi(t)|E(\Delta)|\Psi(t)\rangle
\label{4}
\end{equation}  
holds for all $t$, where $\Delta$ is a Borel set of $R^{3N}$. This equation corresponds to equation (\ref{1}). 

As to Nelson's stochastic mechanics, it is already formulated in the stochastic process language; see for instance \cite{eur}

\vspace{3 mm}
Thus, one can see that the essential elements of a HCT are a set ot trajectories and a measure on it. The Newtonian HCT will be defined in section 5 by giving these two elements. Before this, however, it is necessary to develop some concepts about the structure of the universal wave function. This will be done in the next two sections.

\section{The tree structure of the universal wave function}

The usual representation of the measurement process is the following:
\begin{equation}
\phi\otimes\Phi_0=(\sum_\alpha{c_\alpha \phi_\alpha})\otimes\Phi_0 \rightarrow
\sum_\alpha{c_\alpha\phi_\alpha\otimes\Phi_\alpha},
\end{equation}  \label{6}
where $\phi\otimes\Phi_0$ is the quantum state of the laboratory (composed by microscopical system + measuring apparatus) before the measurement, and $\sum_\alpha{c_\alpha\phi_\alpha\otimes\Phi_\alpha}$ is its state after the measurement. The states $\{\phi_\alpha\}$ of the microscopic system are the eigenstates of the observable which is measured in the experiment, and $\Phi_\alpha$ is the state of the apparatus which has recorded the result $\alpha$ in the measurement.

States $\{\Phi_\alpha\}$ (and therefore also states $\phi_\alpha\otimes\Phi_\alpha$) are spatially disjoined in a permanent way. They are spatially disjoined because they represent an instrument with the pointer in macroscopically different positions, and they are permanently disjoined because of the interaction of the instrument with the environment and the dechoerence process \cite{bohm2}. Actually these states cannot be exactly disjoined, because a wave function with compact support extends immediately over the whole space. In this paper however, it will be assumed that they are exactly disjoined.

\vspace{3 mm}
This situation can be applied to the universal wave function and formalized as follows. The evolution of the universe will be considered only in the time interval $[0,T]$, where $T$ is a very large but finite time. During its evolution, the universal wave function is subjected to many splitting of the above described kind. Thus, at every time $t$, it can be univocally expressed as the sum of permanently disjoined irreducible (PDI) wave packets, i.e.
\begin{equation}
\Psi(t)=\Phi_1+\ldots+\Phi_n,
\end{equation}  \label{7}
where $U(s)\Phi_i$ and $U(s)\Phi_j$ have disjoined support for $i\neq j$ and $0\leq s\leq (T-t)$, and no $\Phi_i$ can be expressed as the sum of permanently disjoined wave packets. Two distinct PDI wave packets $\Phi_i$ and $\Phi_j$ correspond to different macroscopical configurations, while two space points belonging to the support of the same PDI wave packet are macroscopically indistinguishable. Without loss of generality (as it will be more clear in the next sections), we can assume that $\Psi_0$ is composed by only one PDI wave packet. One can see that the universal wave function has a tree structure, with branches generated by the splitting of the PDI wave packets.

\vspace{3 mm}
A subset $S\subset R^{3N}$ is said a {\it PDI-support} if it is the support of a PDI wave packet at a certain time $t$. Let ${\cal S}(t)$ denote the set of the PDI-supports at time $t$, and ${\cal S}$ the set of all PDI-supports. We will assume that all PDI-support are Borel sets. If $S_1\in {\cal S}(t_1)$ is the support of a wave packet $\Phi$, let $S_1 (t_2)$ denote the support of $U(t_2-t_1)\Phi$, where $t_2\geq t_1$ (note that $S_1(t_2)$ may not be a PDI-support). The following equation holds:
\begin{equation}
E[S_1 (t_2)]\Psi(t_2)=U(t_2-t_1)E(S_1)\Psi(t_1).
\end{equation} 
If $S_1\in {\cal S}(t_1)$ and $S_2\in {\cal S}(t_2)$ are two PDI-supports, with $t_1\leq t_2$, and $S_1(t_2)\subseteq S_2$, we say that $S_1\leq S_2$, and this relation is a partial ordering on ${\cal S}$. If $S_1\not \leq S_2$ then $S_1(t_2)\cap S_2=\emptyset$. Moreover, if $S_1,S_2\leq S_3$, we have either $S_1\leq S_2$ or $S_2\leq S_1$.

If $S_1\in {\cal S}(t_1), \ldots, S_n\in {\cal S}(t_n)$ are $n$ PDI-supports, with $t_1\leq\ldots\leq  t_n$, then
\begin{equation}
E(S_n)U(t_n-t_{n-1})E(S_{n-1}) \ldots 
U(t_2-t_1)E(S_1)\Psi(t_1)=E(S_n)\Psi(t_n)
\label{8}
\end{equation}
if $S_1\leq\ldots\leq S_n$, and 0 otherwise.

\section{The classical structure of the branches}
In this section the following conjecture will be discussed:

\newtheorem{assumption}{Conjecture}
\begin{assumption} \rm 
Let $S_1\in{\cal S}(t_1)$, $S_2\in{\cal S}(t_2)$ and $S_3\in{\cal S}(t_3)$ be three PDI-supports with $t_1<t_2< t_3$ and $S_1\leq S_2\leq S_3$. If $\lambda_c$ is a classical (i.e. Newtonian) trajectory so that $\lambda_c(t_1) \in S_1$ and $\lambda_c(t_3) \in S_3$, then $\lambda_c(t_2)\in S_2$.
\end{assumption}
In other words, classical paths go along the branches of the universal wave function. A justification, not a rigorous proof, will be given for this conjecture.

\vspace{3mm}
By posing $\lambda(t_1)=x_1$ and $\lambda(t_3)=x_3$, due to equation (\ref{8}) we have that
\begin{equation}
\langle x_3 | U(t_3-t_1)E(S_1)|\Psi(t_1)\rangle=\langle x_3 | U(t_3-t_2)E(S_2)U(t_2-t_1)E(S_1)|\Psi(t_1)\rangle,
\end{equation}  
i.e.
\begin{equation}
\int_{S_1}{K(t_3,x_3,t_1,x)\Psi(t_1,x)dx}=\int_{S_1}{K'(t_3,x_3,t_1,x_1) \Psi(t_1,x)dx},
\label{10}
\end{equation}
where $K(t_3,x_3,t_1,x)$ is the Feynman propagator obtained by summing over all the paths $\lambda$ for which $\lambda(t_1)=x$ and $\lambda(t_3)=x_3$, while $K'(t_3,x_3,t_1,x)$ is the propagator obtained by summing over all the paths with the previous constraints plus the constraint that $\lambda(t_2)\in S_2$. From (\ref{10}) we can assume that
\begin{equation}
K(t_3,x_3,t_1,x_1)=K'(t_3,x_3,t_1,x_1).
\end{equation}  
Because of the main contribution to the path integral comes from the paths which are close to the classical one, it is reasonable to assume that the classical path $\lambda_c$ which joins $(t_1,x_1)$ and $(t_3,x_3)$ contributes to $ K'(t_3,x_3,t_1,x_1)$, i.e. that it satisfies the constraint $\lambda_c(t_2)\in S_2$. q.e.d.

\vspace{3mm}
Note that the classicality of the branches derives from the fact that the paths which are close to the classical path give the main contribution to the Feynman propagator. In section 7, the same criterion will be used in order to define trajectories for spinning particles, also if no classical counterpart does exist.

\section{A Newtonian HCT}

If Conjecture 1 is correct, it is easy to built a Newtonian HCT. Let $S_0$ be the support of $\Psi_0$, which has been assumed to be itself a PDI wave packet. Let us define
$$
\Lambda:=\{\lambda_c:[0,T]\rightarrow R^{3N} | \lambda(0)\in S_0\},
$$
\begin{equation}
{\cal F}:=\{z^{-1}_T(\Delta)|\Delta\in {\cal B}\},
\end{equation}  
$$
\mu(\Sigma):=\parallel E[z_T(\sigma)]\Psi(T) \parallel ^2,
$$
where $\{\lambda_c\}$ are classical trajectories, and 
$\sigma\in {\cal F}$. In other words, $\Lambda$ is the set of classical trajectories whose initial position is in $S_0$, and the measure on them is defined by applying the quantum measure $||E(\Delta)\Psi(T)||^2$ to their positions at the time $T$.

With this definition we have that
\begin{equation}
\mu_t(S_t)=\langle\Psi(t)|E(S_t)|\Psi(t)\rangle,
\label{15}
\end{equation}  
where $S_t\in {\cal S}(t)$. Indeed, from the conjecture 1 the following lemma can be derived:
\newtheorem{lemma}{Lemma}
\begin{lemma} \rm
Let $S_1,S'_1\in{\cal S}(t_1)$, where of course $S_1\cap S'_1=\emptyset$, and let $t_2>t_1$. Then:

\vspace{1 mm}
(a) $S_1(t_2)\subseteq z_{t_2}[z^{-1}_{t_1}(S_1)]$;

\vspace{1 mm}
(b) $S'_1(t_2)\cap z_{t_2}[z^{-1}_{t_1}(S_1)]=\emptyset.$
\end{lemma}
We have the following picture: by definition, all the trajectories begin inside the root of the tree, and they follows the branches. A set (of zero measure) of them can go out of the branches, but they cannot go into other branches; moreover, directly from conjecture 1, it  follows that the trajectories which go out from a branch cannot go again into the same branch.

From the lemma it follows that
$$
\parallel E\{z_{T}[z^{-1}_t(S_t)]\}\Psi(T) \parallel ^2=
\parallel E[S_t(T)]\Psi(T) \parallel ^2.
$$
Thus:
$$
\mu_t(S_t)=
\mu[z^{-1}_t (S_t)]=
\parallel E\{z_T[z^{-1}_t (S_t)]\}\Psi(T) \parallel ^2=
\parallel E[S_t(T)]\Psi(T) \parallel ^2=
\parallel E(S_t)\Psi(t) \parallel ^2.
$$

Equation (\ref{15}) differs from equation (\ref{4}) because it does not hold for all subset $\Delta$ but only for these subsets which are PDI-support. Inside a PDI support, $\mu_t$ could be a very strange and twisted measure. However this is not important, because the points inside a PDI-support are indistinguishable. For this reason, also equation (\ref{15}) give rise to a HCT which is empirically indistinguishable from quatum mechanics.

\vspace{3mm}
A remark must be made about the $\sigma$-algebra ${\cal F}$ for $\Lambda$. Actually, due to the fact that more trajectories can have the same final point $\lambda_c(T)$ there is no guarantee that the map $z_t$ is measurable, i.e. that for all $t$ and for all $\Delta\in {\cal B}$ we have $z^{-1}_t(\Delta) \in {\cal F}$. This problem can be overcome simply extending the $\sigma$-algebra ${\cal F}$ to the $\sigma$-algebra ${\cal F}'$ generated by the sets $z^{-1}_t(\Delta)$. But the extension of the measure $\mu$ to ${\cal F}'$ is not unique, so it could be possible to have different definitions for the observable measures $\mu[z^{-1}_t(S_t)]$. However this is not the case. Indeed, let us suppose that the set
$z^{-1}_t(S_t) \not \in {\cal F}$; but $z^{-1}_t(S_t)= z^{-1}_T[S_t(T)] \cup \{z^{-1}_t(S_t) \setminus z^{-1}_T[S_t(T)]\}$, where $z^{-1}_T[S_t(T)]\in {\cal F}$, and $\{z^{-1}_t(S_t) \setminus z^{-1}_T[S_t(T)]\}$ is contained in a  ${\cal F}$-measurable set of 0 measure. As a consequence, every extension of $\mu$ to ${\cal F}'$ will give 0 measure for the set $\{z^{-1}_t(S_t) \setminus z^{-1}_T[S_t(T)]\}$, and therefore it will give the same measure for the set $z^{-1}_t(S_t)$.

\section{Physical interpretation}

How is it possible that classical trajectories reproduce quantum phenomena such as, for instance, the interference phenomena of the two-slit experiment? The answer is that a Newtonian HCT violates a propriety which seems so obvious that it is always implicitly assumed in the analysis of conceptual experiments such as the two-slit experiment or the EPR paradox. It is the so called {\it independence assumption} (IA), introduced and widely discussed by Price \cite{price1}\cite{price2}; one of its many possible formulations is: 
\begin{quote} 
{\it The statistical distribution of the variables (direction, energy, hidden variables,...) of the particles emitted by a source does not depend on the future interactions of the particles.}
\end{quote}

IA is violated by the above mentioned Newtonian HCT because the measure on the trajectories is not defined on their initial conditions, but rather on their final conditions. Let us now show how the interference phenomena in the two-slit experiment can be explained in the context of the Newtonian HCT.

Let us consider the version of the experiment performed with electrons:
\begin{center}
\unitlength=1mm
\begin{picture}(120,35)
\put(54,5){\line(1,0){10}}
\put(54,30){\line(1,0){10}}
\put(58,17){\circle*{1}}
\put(112,5){\line(0,1){23}}
\put(0,16){\line(1,0){7}}
\put(7,16){\line(0,1){3}}
\put(0,19){\line(1,0){7}}
\put(0,16){\line(0,1){3}}
\bezier{600}(7,18)(58,22)(112,17)
\bezier{600}(7,17)(58,13)(112,17)

\put(47,2){\makebox(6,6){$E_2$}}
\put(47,27){\makebox(6,6){$E_1$}}
\put(50,14){\makebox(6,6){$F$}}
\put(105,1){\makebox(6,6){$H$}}

\put(0,10){\makebox(6,6){$S$}}

\end{picture}

Fig. 1
\end{center}

Here $S$ is an electron source, $F$ is a tiny conducting wire which crosses the plane of the figure at right angle and can be set to a positive potential with respect to the two electrodes $E_1$ and $E_2$; $H$ is a screen constituted by a photographic plate. Due to the electrostatic field generated by the wire, the electrons emitted by the source are deflected and produce interference fringes on the screen. If the electrostatic field is turned off, the interference fringes disappear.

In a description of the motion with classical trajectories, the point of the screen hit by an electron depends on the little angle with which the electron is emitted by the source, and the figure on the screen depends on the statistical distribution of this angle. Maintaining the trajectory description, one can explain the interference fringes by accepting that the statistical distribution of this angle changes when the electrostatic field is turned on and off, even if no interaction exists between the source and the deflecting device. This is the violation of IA, which, in the context of a Newtonian HCT, is not due to strange backward interactions, but simply to the structure of the measure $\mu$.

\vspace{3 mm}
The IA is also assumed in the analysis of the EPR paradox \cite{price1}, so that there is no reason for which a suitable Newtonian HCT which includes spin particles cannot reproduce that kind of experimental results.
\section{Conclusion}
The results of the present paper are the following.

(1) It has been shown that a hidden-configuration theory (HCT) can expressed in the mathematical language of stochastic processes, i.e. by defining a set of trajectories $\lambda:R^+\rightarrow R^{3N}$ with a measure defined on it.

(2) Starting from the analsys of the measurement process, it has been shown that the universal wave function has a tree structure, where the branches of the tree have disjoined support.

(3) It has been conjectured that the branches of the tree have a classical structure, i.e. classical trajectories go along them. This feature derives from the fact that classical paths give the main contribution to the quantum propagator.

(4) a Newtonian HCT --i.e. a HCT with classical trajectories-- has been proposed, which could be empirically equivalent to quantum mechanics.

(5) Quantum phenomena, like for instance the interference effects of the two-slit experiment, are explained in the context of a Newtonian HCT by assuming the violation of the Independence Assumption, i.e. by assuming that the distribution of the particles emitted by a source can depend also on the future interactions of the particles. This violation however has not a dynamical origin, but it depends on the measure, in particular by the fact that the measure is not defined on initial conditions of the trajectories but on their final positions.


\end{document}